# Liquid water based optoelectronic computing chip

*Minhui Yang[1], Kangchen Xiong[1], Xin Chen[1], Huikai Zhong[1], Shisheng Lin,[1,2]\**

[1]*College of Information Science and Electronic Engineering, Zhejiang University, Hangzhou, 310027, P. R. China.*

[2]*State Key Laboratory of Modern Optical Instrumentation, Zhejiang University, Hangzhou, 310027, P. R. China.*

## Abstract

Solid-state electronics have revolutionized modern society due to their exceptional computational capabilities. However, the power consumption of chips rises dramatically with increasing integration levels as post-treatment of individual computing unit cost. Here, we introduce a water computing chip with exceptionally low power consumption based on graphene/water/semiconductor photodetectors, which relies on a unique exponential decay of water molecule polarization transfer function (PTF) along the signal transmission direction. Through a designed Ising model to picture the dynamic polarization and depolarization process of water molecules between graphene and silicon, we illustrate a longitude decay PTF which guarantees low energy consumption for the pulse current output. Furthermore, the lateral decay PTF is verified by the realization of linearly superimposed currents in the diameter of centimeter scale based on 3×3 water photodetector array. The molecular dynamic simulations reveal the timescale of 25 femtosecond (fs) for one flip of single water molecule. Subsequently, the single logical operation only needs an ideal energy consumption up to attojoule ($10^{-18}$ J). The decay of lateral PTF at the centimeter scale ensures the ability to fabricate water computing chips containing a large number of photodetector arrays under micro/nano fabrication technology, which mimics the numerous neural cells inside hippocampus. As a proof of concept, we show the capability for Multiply-Accumulate (MAC) computations using an 8×8 water computing chip by successfully identifying the ASCII code of 'ZJU'. The water computing chips utilize naturally-sourced water and straightforward

**manufacturing techniques, which hold great potential in creating a chip similar with human brain.**

## Introduction

The evolution of solid-state chips is accelerating as the computing power demand of solid-state chips is seriously challenged by the rapid development of artificial intelligence (AI)[1-5]. With the highly distributed neurons which can process information parallelly in AI network, sophisticated applications that require huge computation including real-time multi-target detection and classification can be easily realized by improving the quantity of the neuron[6-11]. As of now, most of the neural network chips mimic the function of spiking neurons through the use of analog computations, where the connection between the neurons is conducted by electronic circuits governed by differential equations that mirror the characterization of the membrane potential in biological neurons[12-14]. Up to now, the lowest energy cost of electronic neural network chip is about $10^{-12}$ J/OP[15]. In parallel to the electronic neural network chip, the research on optical neural network chips has also received much attention, whose calculation process still relies on optoelectronics conversion, resulting in its energy efficiency on the same order of magnitude as electronic neural network chips[16-18]. Recently, the energy cost of optical neural network chips decreases to 7 femtojoules (fJ) per operation through utilizing one type of coherent optical neural network, while its post-treatment of individual computing unit is still dependent on traditional circuit components[15].

In parallel with the extensive works on the development of more versatile chips, the further miniaturization of chips has also made tremendous progress. The chip unit has now reached the level of 5 nm[19-21], greatly increasing the complexity of manufacture process and posing a huge challenge to Moore's Law[22-24]. Besides, the power consumption of the chip rises with increasing integration levels which in turn presents a challenge to the demand for further enhancement of computational power in the era of big data. Herein, a new way to reduce the number of components required to achieve the neural network calculation is highly demanded. As one of the most common substances in nature, water is polar owing to the asymmetrical polar covalent bonds, which is composed of two hydrogen atoms and one oxygen atom. The polarity of water molecules allows liquid water hold great potential in the application in life science, energy, materials and other fields[25-28]. It has been proved that water molecules have an ultrafast vibrational dynamic with less than 100 fs for intermolecular vibrations and 10 fs for intramolecular vibrations[29-31]. Besides, the pulse current response we observed in water-based photodetectors shows the great potential of water-based optical devices in low energy consumption computing with high speed as the water molecules undergo a mechanical polarization flip excited by light[32-35].

Here, we propose a novel water computing chip which is capable of self-actuating linear superposition calculations based on polarization transfer function (PTF). We utilize a designed Ising model embedded by Transferable Intermolecular Potential 3-point (TIP3P) to depict the polarization-depolarization process in a graphene/water/semiconductor photodetector, which is the unit of water computing chip, predicting the existence of longitude PTF in water molecules. The rapid decay characterization of longitude PTF ensures the pulse-like form of photocurrent, leading to a low energy cost. Besides, we construct a protype of 3×3 water computing chip which can perform linear superposition calculation to demonstrate the lateral PTF can reach centimeter magnitude. This centimeter scale lateral PTF decay in the hippocampus comparable scale enables the water computing chip to integrate larger scale photodetector arrays under micro/nano fabrication technology. In addition, molecular dynamics simulations demonstrate the speed of polarization flip of water molecules is less than 25 fs, showcasing the substantial computing capability of water computing chips with an energy cost down to $10^{-18}$ J/OP. Furthermore, we accurately identify the ASCII code of 'ZJU' through utilizing an 8×8 water computing chip. Compared with solid-state chips, the connection between the photodetectors in the water computing chip can be established directly and naturally through an exponential decay PTF. The introduction of the water computing chip holds significant implications for modeling brain computation patterns.

## Results

**Structural and optoelectronic response characterization.** The schematic illustration of the graphene/water/silicon photodetector is depicted in Fig. 1a, where water molecules are positioned between graphene and n-type silicon. The molecular picture of the device is shown in Fig. 1b, where the majority of water molecules are in a disordered phase without the light illumination. The silicon was n-type with a resistivity of 5 Ω cm. Additionally, deionized (DI) water was prepared by ultra-pure water system with a specific conductance < 20.1 μS cm$^{-1}$. In order to measure the optical response when we light on the graphene layer, a Keithley 6514 source meter was connected to the two electrodes, where the positive end was connected to the Au electrodes. Fig. 1e, f show the response photocurrent with DI water of room temperature and freezing temperature under 500 nm illumination respectively. As we can see, pulse-like photo-generated current can be produced at room temperature while the liquid photodetector loses its response when water is frozen into ice. We deduce that the water molecules play an important role for the generation of pulse-like response current. When the temperature is above the freezing point, the Fermi level difference between graphene and silicon can induce the polarization of water molecules, causing the water molecules to flip at the interface of graphene/water and silicon/water (oxygen atoms toward graphene, hydrogen atoms toward silicon) as seen in Fig. 1c. Subsequently, the light exciting the electrons in the silicon and graphene will induce more water molecules polarized at the water/silicon interface and water/graphene interface. The polarized water molecules will further facilitate the transfer of electrons from silicon to graphene. There is no doubt that the mobility of liquid is essential for water molecules to be polarized. However, ice exhibits a typical hexagonal lattice with strong hydrogen bonds between oxygens and hydrogens as shown in Fig. 1d, which extremely restrain the polarization flip of water molecules, which the disappearance of pulse-like photo-generated current in graphene/ice/silicon photodetector further verify the phenomenon of water molecule polarization in liquid heterojunctions.

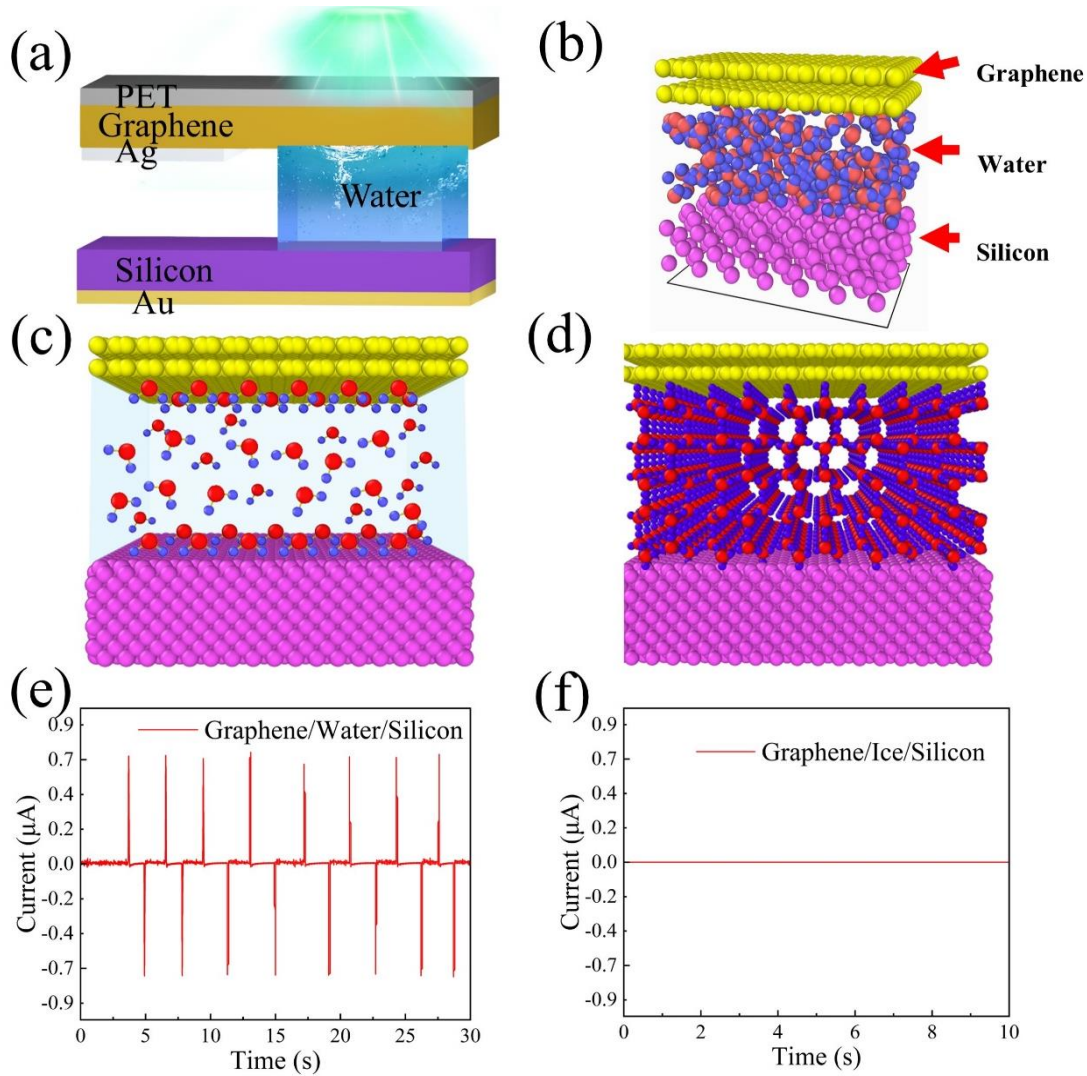

**Fig. 1 | The structure of the graphene/water/silicon photodetector and its response characteristics a,** Schematic diagram of the device structure of the self-driven photodetectors. **b,** Diagram of the graphene/water/silicon molecular structure without light illumination. **c,** Diagram of the graphene/water/silicon molecular structure under 500 nm illumination. **d,** Diagram of the graphene/water/silicon molecular structure when water is frozen into ice. **e,** Time dependent photocurrent curve of graphene/water/silicon photodetector. **f,** Time dependent photocurrent curve of graphene/ice/silicon photodetector.

**Ising model embedded by TIP3P.** Considering that similarity between the polarization flip of water molecules and the electron spin state, a designed Ising model was employed to illustrate the dynamic transition of water molecules from disordered phase to polarized phase. The Ising model is widely studied in ferromagnetic systems[36-37]. It describes the behavior of numerous magnetic spins, arranged on a lattice, capable of taking values of +1 (up) or -1 (down). The spins interact with each other through Hamiltonian that depends on the orientation of neighboring spins, as determined by the following formula (1).

$$H = -J \sum_{<i,j>} s_i K_{ij} s_j \qquad (1)$$

where $s$ is the spin state, which is set as $s \in \{-1,1\}$, the coupling matrix $K_{ij}$ is a real symmetric matrix which describes the interaction strength between $s_i$ and $s_j$. $<i,j>$ is the combination of adjacent spins, $J$ is energy coupling coefficient. The flip of spin states is governed by the Gibbs distribution, as determined by formula (2).

$$p(s) = \min(e^{-\frac{H(s)}{k_B T}}, 1) \qquad (2)$$

where $k_B$ is the Boltzmann constant, $T$ is the thermodynamic temperature. Over time, the Ising system tends to transition gradually from disorder to order.

The state of water molecules within the Ising lattice is denoted by spin state. Given that silicon attracts positively charged hydrogen atoms, and graphene attracts negatively charged oxygen atoms, a water molecule with +1 spin implies oxygen atoms face up, while hydrogen atoms face down symmetrically. The TIP3P model is employed to calculate the coupling matrix $K_{ij}$. The TIP3P model, is a widely utilized representation of water in molecular simulations. The model comprises three individual atoms—two hydrogen atoms and one oxygen atom—each assigned specific charges. In the TIP3P model, the hydrogen atoms are positioned at a distance of 0.9572 angstrom from the oxygen atom, and the bond angle of H-O-H is precisely 104.52°. The hydrogen and oxygen atoms carry partial charges of +0.417 and -0.834, respectively, reflecting the electronegativity differences between the two elements. The TIP3P model has accurate reproduced various properties of real water, encompassing structural and thermodynamic characteristics. Through TIP3P model, the coupling matrix $K_{ij}$ can be determined by the following formula (3).

$$K_{ij} = \sum_m \sum_n \frac{k_c q_m q_n}{r_{mn}} + \frac{A}{r_{oo}^{12}} - \frac{B}{r_{oo}^{6}} \qquad (3)$$

where $k_C$ is the electrostatic constant, $A$ and $B$ are the Lennard-Jones parameters, $m$ and $n$ are the atomic number within different water molecules, $r_{ij}$ is the distance between two atoms, $q_i$ and $q_j$ are the charge of hydrogen or oxygen atoms, $r_{OO}$ is the distance between two oxygen atoms. The formula for $K_{ij}$ is composed of two distinct components: the coulomb potential and the Lennard-Jones potential. The Coulomb potential involves the summation of all interacting atoms between two water molecules, while the Lennard-Jones potential is exclusively dependent on the interaction between oxygen atoms. Besides, the Ising lattice restricts water molecules to two states—up or down—resulting in six possible relative orientations: front, back, top, bottom, left, and right. The oxygen atom is consistently positioned at the lattice point, where $r_{oo}$ is equivalent to the Ising lattice constant. These analyses allow us to compute the interactions between water molecules in the Ising lattice for any given scenario.

After obtaining the coupling matrix, it is crucial to consider more realistic simulation conditions. Due to the Fermi level difference between graphene and n-type silicon, water molecules on the external layer will be initially polarized. Under illumination, incident photons excite electrons in the silicon and graphene, leading to enhanced polarization of water molecules. To explain this phenomenon in the Ising model, we construct ten layers of Ising lattice. The Fermi level difference between graphene and n-type silicon indicates a binding effect on the polarization state of surface water molecules. Due to the binding effect, the spins of surface water molecules do not flip during the evolution of the Ising model. The transition of inner water molecules from unpolarized phase to polarized phase is illustrated by the transfer ability of polarization which presents an exponential decay as quantified by formula (4).

$$P = Ae^{-\tau \cdot d^{\alpha}} \tag{4}$$

where $P$ is the probability of inner water molecules being polarized, $A$ is the coefficient of polarization, $\tau$ is the polarization transfer coefficient, $d$ is the distance from the surface, $\alpha$ is the interlayer differentiation factor. Formula (4) indicates that the rate of polarization decreases as the water molecules moves farther away from the surface layer which successfully predicts the existence of longitudinal PTF.

Additionally, the spins of the system tend to become ordered or disordered corresponding to the presence or absence of light. The tendency of spin become disordered attributed to the depolarization rate exceeds the polarization rate when the light is turned off. In accordance with the energy function of the Ising model, when J > 0, the energy of the adjacent node is minimized in the condition of all the spins aligned in the same direction. On the contrary, the system energy increases when these nodes are misaligned.    As a result, the sign of $J$ is flipped when the light is turned on or off. Subsequently, we employ Monte Carlo methods to simulate the evolution of water molecules in the Ising model embedded by TIP3P. The simulation involves the following steps using Monte Carlo methods: In the initial step, a lattice comprising ten layers is established, with each grid point representing a water molecule. Subsequently, a spin is randomly selected and its orientation is altered from up to down. The coupling matrix $K_{ij}$ between the chosen spin and adjacent spins is then calculated using Formula (3). Following this, the energy of the new spin configuration is calculated by Formula (1). The new spin configuration will be accepted if the energy become lower; otherwise, it is accepted with a probability as delineated in Formula (2). This process is repeated until the predetermined time step N is reached. Significantly, the energy function curve, shown in Fig. 2a, demonstrates a sudden change when the light is turned on or off, leading to an ultrafast current response. Fig. 2b-e visually depict the process of water molecules polarization, with green representing +1 spin, red denoting -1 spin of the water molecule and black indicating no polarization of water molecules.

Ising model vividly simulates the polarization-depolarization process of water molecules, and predicts that the polarization of water molecules has the transfer ability which is called PTF. PTF in Ising model is longitudinal which ensures the low energy consumption for the pulse current output. The discovery of PTF with exponential decay promotes the construction of novel water computing chips on the basis of the unique PTF of water molecules as the medium of electrical signal transmission.

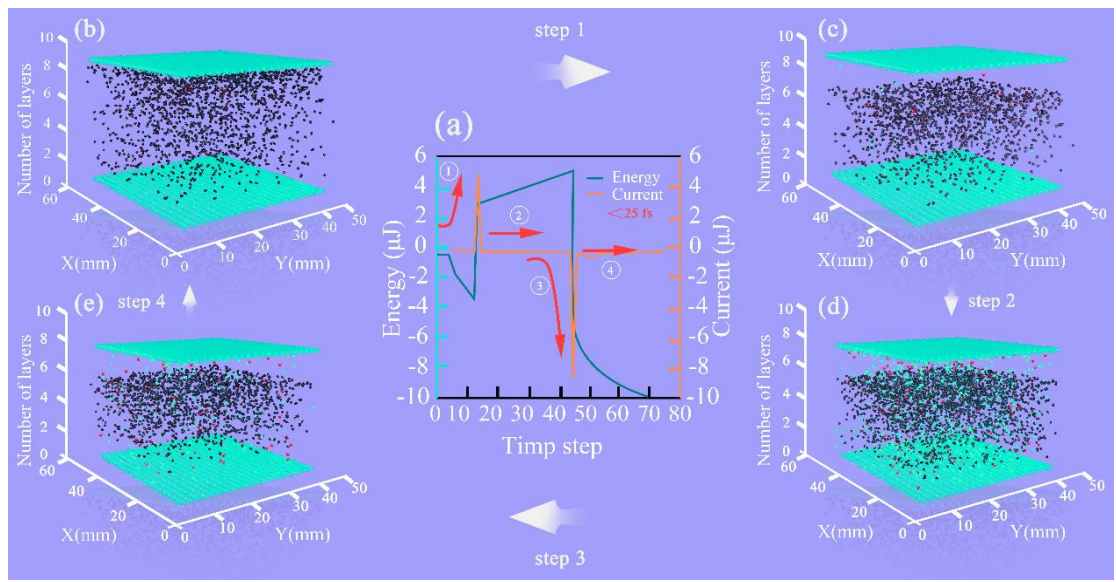

**Fig. 2 | Results of polarized water molecules simulated by Ising model embedding by TIP3P. a,** The energy curve and the current curve simulated by Ising model embedding by TIP3P. **b,** Initial state of the graphene/water/silicon photodetector. **c,** Polarization of surface water molecules induced by light. **d,** Polarized water molecules

transfer to the deep layers under continuous illumination. **e,** Depolarization process of water molecules without illumination.

**PTF of water molecules.** Ising model has successfully demonstrated that the polarization of water molecules has transfer ability by introducing PTF with exponential decay. It is necessary to emphasize the core of water computing chip is the PTF which will establish the signal connection between each photodetector naturally and directly, leading to the self-driven inference process ability of the water computing chip. Fig. 3a pictures the propagation process of lateral PTF. The lateral PTF of DI water at various distances is depicted in Fig. 3b, clearly exhibiting exponential decay as the distance increases. Specifically, it reaches half of its maximum value at 1 cm and decreases to 20% of the maximum value at 50 mm. Generally, the attenuation of PTF follows a fast-then-slow trend. The lateral PTF decay at the centimeter scale provides the water computing chip with the capability to accommodate a large number of photodetector arrays under micro/nano fabrication technology. When various glucose concentrations ranging from 10 g/L to 160 g/L are introduced into water, PTF exhibits exponential decay, as illustrated in Fig. 3c. This observation affirms that PTF can be modulated by different solutes in DI water. At a glucose concentration of 10 g/L, the current reaches its minimum, measuring 0.38 nA. Conversely, at a concentration of 25 g/L, the current achieves its maximum, reaching 0.16 µA. Subsequently, the current diminishes with increasing concentration, and a plateau is observed between 40 g/L and 60 g/L, with a value of 0.13 µA. In Fig 3d, despite a reduction in the light-receiving graphene area from 0.29 cm² to 0.09 cm², the lateral PTF still exhibits exponential decay, while the maximum current at 1 mm decreases from 0.40 µA to 0.23 µA. The inset of Fig 3d clearly illustrates the current reduction as the light-receiving graphene area decreases at 10 mm PTF. The preservation of the exponential decay property despite area reduction indicates that even a small number of water molecules can excite PTF.

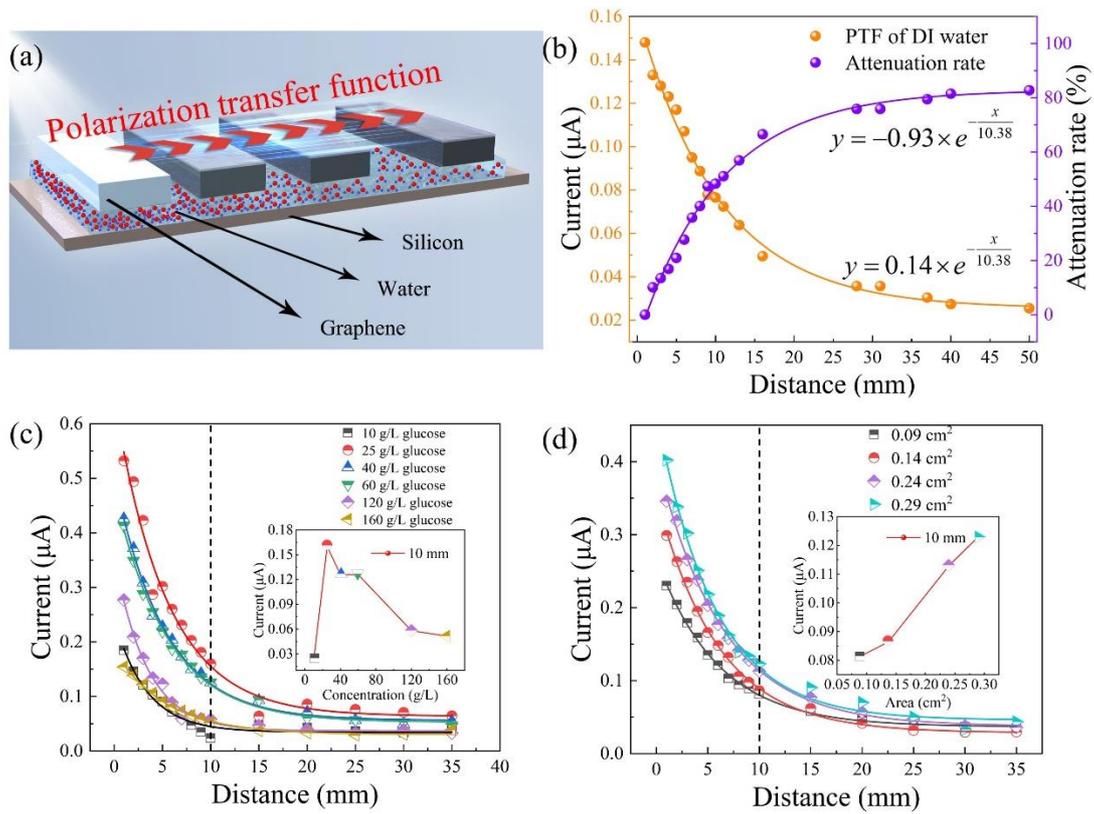

**Fig. 3 | Lateral PTF under different conditions. a,** Physical images and experimental setup of PTF of water molecules. **b,** PTF of DI water and its attenuation rate. **c,** PTF of water with glucose concentrations of 10 g/L, 25 g/L, 40 g/L, 60 g/L, 120 g/L, 160 g/L. The inset shows the current of each PTF at 10mm. **d,** PTF of light-receiving graphene areas of 0.09 cm$^2$, 0.14 cm$^2$, 0.24 cm$^2$, 0.29 cm$^2$. The inset shows the current of each PTF at 10mm.

## Discussion

**Linear superposition calculations based on water molecules.** Utilizing PTF, we construct a 3×3 water computing chip as shown in Fig. 4a. As we can see, PTF of water molecules exhibits a clear linear superposition property which enable Multiply Accumulate (MAC). The maximum response currents of the photodetectors at the input layer first multiply with PTF and then superpose at the hidden layer. Subsequently, the current at the hidden layer enhance corresponding to the light input where '0' represents that the detector is unilluminated while '1' means that the detector is illuminated. Finally, the output layer obtains the signal transmitted from the hidden layer through PTF. Fig. 4d visualizes the current response of the photodetector positioned in the middle of the output layer, given light input to the hidden layer of (1 0 1). Fig. 4b, c report the current responses from photodetectors positioned in the corner of the output layer and the middle for various combinations of light applied to the hidden layer respectively. The photodetector at the middle of the output layer reaches the maximum current with a hidden layer illumination of (1 1 1) and the minimum current with an illumination of (0 0 0). As the hidden layer illumination shifts from (0 0 0) to (0 0 1), the current exhibits a minor increase. Yet, when the light is (0 1 0), the current rises more than it does at (0 0 1), and with the shift to (0 1 1), the current rises further. In summary, the current generated by the photodetectors in the hidden layer is conveyed to the output layer via the long-distance attenuation PTF, resulting in a reinforcement of the output layer current at a minor amplitude. However, with short-distance attenuation, the output layer current undergoes significant enhancement. In short, a protype of water computing chip, driven by PTF of water molecules, has been constructed, capable of performing linear superpositions.

Notably, our photodetector responds within a single time step of the Ising model. In order to more precisely investigate the response time causing by polarization flip of water molecules, we perform molecular dynamics simulation implemented with LAMMPS, which illustrates that the polarization flip of water molecules occurs on the order of femtoseconds. As depicted in Fig. 4e, f, the time interval of polarization flip is less than 25 fs. In particular, an energy of 2ev (greater than silicon's band gap of 1.12ev) is applied to the silicon during the simulation which ensures electrons in the silicon absorb sufficient energy to transit from the valence band to the conduction band. In conclusion, the rapid flipping speed of water molecules illustrates the substantial computing capability of water computing chips.

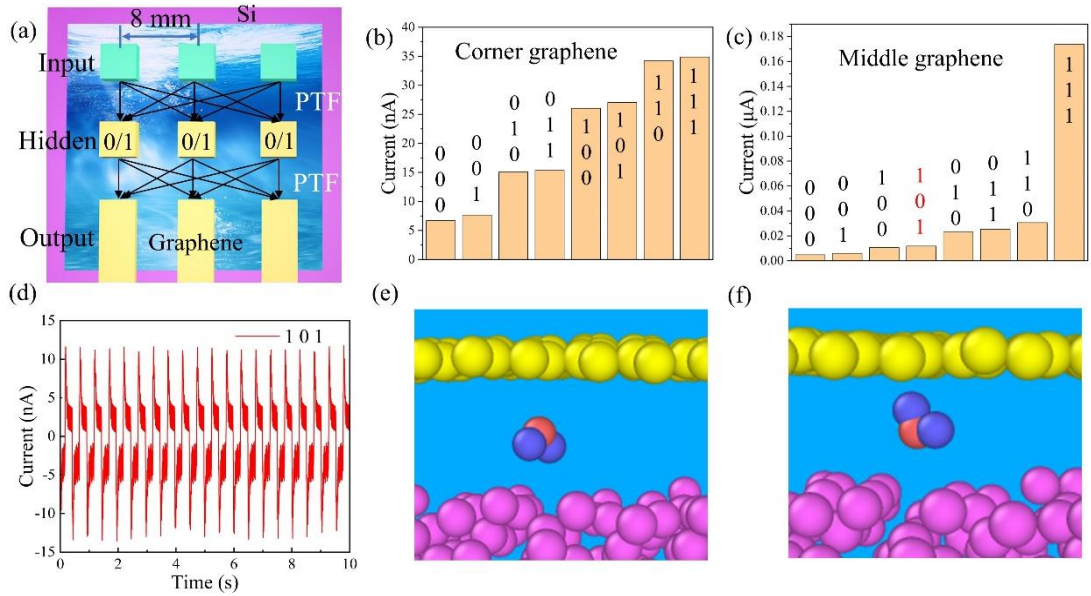

**Fig. 4 | Linear superposition calculations by a 3×3 water computing chip. a,** The 3×3 water computing chip where the spacing between the photodetectors is 8mm. PTFs establish connections between neurons. The photodetectors in the hidden layer are applied with 0/1 illumination **b, c,** Maximum values of the current response of the photodetectors in the corner of the output layer and the middle respectively, corresponding to different combinations of light applied to the hidden layer. **d,** Current response of the photodetector located in the middle of the output layer with (1 0 1) light applied to the hidden layer. **e, f,** Two states of single water molecule which present opposite polarization orientations. The time interval between the two states is less than 25 fs.

**On-chip inference to determine the ASCII code of 'ZJU'.** We constructed a liquid water based optoelectronic computing chip featuring an 8×8 array of photodetectors, which can complete the ASCII code recognition process. In Fig. 5b, the 1$^{st}$ row is the input layer, which maintains the light, the 8$^{th}$ row is the output layer where the magnitude of the current is used to distinguish between 0 and 1 in ASCII code (Fig. 5c) and 2$^{nd}$-7$^{th}$ rows are the hidden layers, which correspond to the tunable LED light field distribution. Four photodetectors framed in the middle of the chip enable MAC, where the currents from the left three photodetectors first multiply with PTF and then accumulate at the next-layer photodetector (Fig. 5a). The recognition procedure involves two stages: online learning and on-chip inference. We designed an algorithm on MATLAB for calculating the optimal LED light field distribution. The mean square error loss function (MSE) between the magnitude of the current at the output layer and the target ASCII code is calculated. If MSE exceeds a threshold, a loop is triggered to optimize the LED light field distribution. The optimal light field distribution in the process of identifying ASCII of "Z" is shown in Fig. 5d. Then the chip completes the inference process under this optimal light field distribution, the magnitude of the current generated by each photodetector shown in Fig. 5e. The accuracy curve for the recognition task is shown in Fig. 5f. After iteration, the water computing chip can accurately recognize the ASCII code.

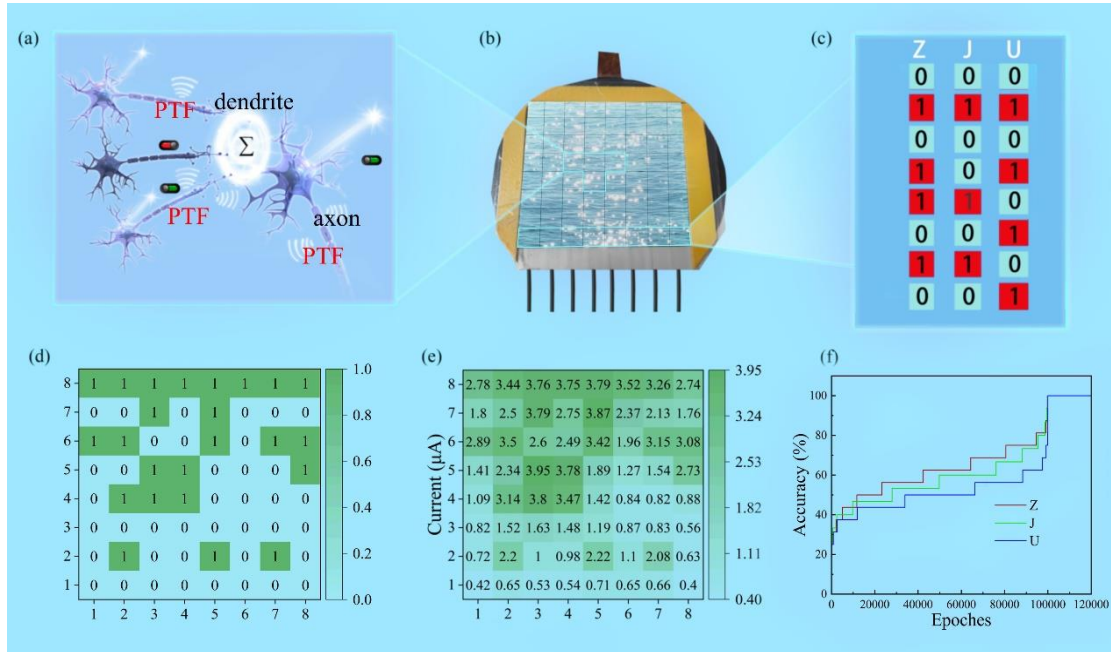

**Fig. 5 | Identification of the ASCII code of 'ZJU' using an 8×8 water computing chip. a,** Schematic representation of neuromorphic calculations for liquid-based photodetector arrays. Connections between photodetectors are constructed through PTF. The signals from pre-neurons are multiplied with PTF and accumulated at the post-neuron. The four neurons constitute the Multiplier and Accumulation (MAC) unit. **b,** Water computing chip composed of 8×8 liquid-based photodetector arrays. **c,** ASCII code corresponding to 'ZJU' used for chip recognition. **d,** The optimal light field distribution for the case with the highest recognition accuracy of "Z". **e,** The magnitude of current generated by each detector on the 8×8 array of photodetectors under the optimal light field distribution. **f** Accuracy curve for ASCII code recognition of 'ZJU'.

**Energy efficiency.** The water computing chip outlined in this paper possesses energy advantages in four key areas. First, the high responsiveness of the photodetector is crucial for converting optical input to an electrical signal. Second, the natural PTF of water molecules enables signal transmission between photodetectors directly, making the chip independent of supplementary components, including ADC/DAC and other energy-intensive devices, during the inference calculation process. Moreover, compared to electronic neural network chip, the liquid-based photodetector does not necessitate electronic circuits governed by differential equations, thereby further minimizing the chip's energy consumption. Additionally, the event-driven, spiking nature of the water computing chip means that photodetectors do not need to maintain light at idle periods. The responsivity of the graphene/water/silicon photodetector is 157.48 mA/W, indicating that 1 µA of current can be produced at an optical power of 6.37 µW. Considering the general case that the light source used by the chip has a wall-plug efficiency of 25%, the power of a MAC unit as shown in Fig. 5a is 101.92 µW. As memory access computes 1 MAC=2 operation (OP), the MAC unit operates at $W=2.55\times10^{-18}$ J/OP, $W=P\cdot T/2$ with $P=101.92$ µW and the width of the pulse current generated by the photodetector $T=50$ fs as the event-driven nature of the water computing chip consumes energy only when light input is available. The power consumption of liquid-based photoelectronic water computing chip is improved by at least two orders of magnitude over solid-based chips[15].

In summary, the water computing chip has brought about a paradigm shift in conventional solid-state chips across multiple dimensions. The inherent PTF of water molecules enables signal transmission between devices, relying solely on natural water molecules. This capability empowers the water computing chip to conduct inference processes independently, without the need for external devices with high energy consumption, thereby significantly reducing the overall energy requirements of chips. Importantly, the transition from solid-state to liquid-based chips creates a working environment that closely simulates the human brain, offering a fresh perspective for the advancement of brain-like computing. It is crucial to emphasize that the decay of lateral PTF is at the centimeter scale which ensures the water computing chips are able to contain a large number of photodetector arrays under micro/nano fabrication technology. Leveraging water, a readily available natural resource holds immense significance for realizing next generation of brain-like computing.

## Methods

**The fabrication of the liquid-based photoelectronic water computing chip.** 1) Using the low-pressure chemical vapor deposition (CVD) method in a quartz tube furnace, monolayer graphene samples were grown on copper substrates. The growth process took place at 1000 °C for 60 minutes, employing $CH_4$ and $H_2$ with a flux ratio of 4:1. Subsequent to the growth, the furnace was gradually cooled to room temperature at a rate of approximately 30 °C/min. 2) Graphene arrays corresponding to the size of the LED arrays are laser etched or mechanically etched on the graphene with polyethylene terephthalate (PET) protection. 3) Electrodes are fabricated on one side of the graphene arrays and the bottom of the semiconductor. 4) Kafuter glue or plastic sheet is applied to the other three sides of the PET film where graphene electrodes are not fabricated. 5) Cover graphene array/PET structures on selected semiconductor. 6) Fill water (1-3mm) between the graphene array and the semiconductor through a syringe or pipette to finalize the fabrication of the liquid-based photoelectronic water computing chip.

**Characterization analysis.** Time-current curves of the liquid-based photodetectors were measured by Keithley 6514 source meter and DMM6500 multimeter. The voltage-current curves of the liquid-based photodetectors were carried out by a Keithley 2400 source meter.

## Conflict of Interest

The authors declare no conflict of interest.